# Systematic Characterization of Hydrophilized Polydimethylsiloxane

Daniel J. O'Brien, Andrew J.H. Sedlack, Pia Bhatia, Christopher J. Jensen, Alberto Quintana and Makarand Paranjape

*Abstract*— Flexible microfluidics have found extensive utility in the biological and biomedical fields. A leading substrate material for compliant devices is polydimethylsiloxane (PDMS). Despite its many advantages, PDMS is inherently hydrophobic and consequently its use in passive (pumpless) microfluidics becomes problematic. To this end, many physical and chemical modifications have been introduced to render PDMS hydrophilic, ranging from amphiphilic molecule additions to surface plasma treatments. However, when transitioning from lab benchtop to realized medical devices, these modifications must exhibit long-term stability. Unfortunately, these modifications are often presented but their mechanisms and long-term stability are not studied in detail. We have investigated an array of PDMS modifications, utilizing contact angle goniometry to study surface energy over a 30-day evolution study. Samples were stored in air and water, and Fourier Transform Infrared-Attenuated Total Reflectance (FTIR-ATR) analysis was used to confirm surface functional group uniformity. We have identified preferred modification techniques for long-lasting PDMS devices and characterized often overlooked material stability.

*Index Terms*—microfluidics, flexible electronics, Biomedical microelectromechanical systems, polydimethylsiloxane,

## I. Introduction

THIS paper is an extension of our previous proceedings article presented at the IEEE MEMS 2020 Conference [1].

The seemingly ubiquitous polymer, polydimethylsiloxane (PDMS), excels in its use in the fields of microfluidics and bioMEMS [2]. PDMS is a relatively inexpensive material, easy to use, simply adaptable to new devices via photolithographic mold fabrication, optically transparent, reproducible, has tunable elastic properties, and is flexible, allowing its use in the biomedical arena, particularly with non-rigid wearable devices [3].

Passive, or pumpless, microfluidics would further expand the horizons of PDMS-based devices for both medical and biological study. However, the material's low surface energy renders it hydrophobic [4]. Many groups have exploited modified surface properties generated by plasma treatment activation [5], polymer grafting [6], bulk surfactant additions [7]–[12], wet chemical treatments [13], and UV/ozone treatments [14]. These approaches introduce polar functional groups to the surface of micro channels, allowing non-driven "autonomous" flow of polar fluids [15]. However, plasma-based approaches are beleaguered by hydrophobic recovery, which occurs due to diffusive molecular rearrangement of low-molecular weight chains (arising from polymer chains that did not substantially crosslink) from the bulk to replace thermodynamically unstable surface moieties [16]. The stability and mechanisms of these modifications are often overlooked in favor of rapid publication and applications.

Our group is designing a microfluidic device for in-situ, lab-on-a-chip biomarker sensing in passively sampled interstitial fluid (ISF) [17]. This endeavor requires a complex, multifunctional device with several fabrication steps, underscoring the need for simplicity of the materials and processes involved. Additionally, the device fails with un-modified PDMS, as its hydrophobic character prevents capillary pressure-driven fluid transport (contact angle ~115°). Accordingly, we have compiled and tested a diverse suite of simple techniques for PDMS hydrophilization. The studied modifications include (1) dynamic surface modification through surfactant addition to the PDMS prepolymer, (2) substrate roughening for microstructuring of the PDMS surface, (3) nanostructuring of PDMS surfaces via reactive ion etching, (4) $O_2$ plasma silanolization (replacing surface methyls with silanols), (4) low-molecular weight chain extraction, (5) surface grafting of a hydrophilic polymer, and (6) reactive magnetron oxide sputtering. Additionally, we tested the effect of two commonly used PDMS additives, benzophenone and carbon black, on hydrophilization. Benzophenone is used to allow photo-patterning of PDMS, in place of traditional mold setting [18]. Carbon black is a pigment used to generate conductive PDMS composites [19], gas selective membranes [20], and for light absorption to allow optical segregation of device components/layers [21], [22].

A systematic approach to characterize long-term hydrophilic stability was undertaken, using static contact angle measurements over a thirty-day period, with measurements conducted in an isolated environment. Temperature and relative humidity were controlled so as to limit the variability in surface hydrophilicity, as observed in literature [23], [24]. Mechanisms of hydrophilization were characterized via surface functionality using infrared spectroscopy (FTIR-ATR), surface roughness

This paper was first submitted for review on May 6, 2020. The authors thank the Achievement Rewards for College Scientists-Metro Washington Chapter (ARCS-MWC) and the Forster Family Foundation, as well as National Science Foundation for financial assistance through Award #1938995. We acknowledge Leon Der and Jasper Nijdam for technical assistance, as well as thank GNuLab and ISM² at Georgetown. Oxide depositions were performed in the Kai Liu Lab at Georgetown.

All authors are associated with Georgetown University, Washington, DC 20057 USA. (e-mails: djo44@georgetown.edu, paranjam@georgetown.edu).

TABLE I
SURFACTANTS & CONCENTRATIONS (WT. %)

| SILWET L-77 | TRITON X-100 | TWEEN 20 | PDMS-B-PEO | PLURONIC F-127 (200 mg/mL EtOH) |
|---|---|---|---|---|
| 0.2 % | 1% | 1% | 0.5% | 2 µL/g PDMS |
| 0.4 % | 2% | 2% | 1.0 % | 6 µL/g PDMS |
| 0.6 % | 3% | 3% | 1.5 % | 10 µL/g PDMS |

TABLE II
RIE RECIPES FOR NANOSTRUCTURING OF PDMS SURFACES

| RECIPE | STEP | GAS | FLOW [SCCM] | TIME [MIN] | PRESSURE [MTORR] | POWER [W] |
|---|---|---|---|---|---|---|
| 1 | 1 | $O_2$ | 20 | 15 | 75 | 150 |
| 2 | 1 | $O_2$ | 20 | 30 | 75 | 150 |
| 3 | 1 | $SF_6$ | 60 | 5 | 30 | 600 (ICP) |
|   | 2 | $O_2$ | 20 | 1 | 100 | 100 |
| 4 | 1 | $SF_6$ | 60 | 10 | 30 | 600 (ICP) |
|   | 2 | $O_2$ | 20 | 1 | 100 | 100 |

using atomic force microscopy (AFM), and data modeling to extract surfactant diffusivities through the thermoset PDMS matrix. This wide-ranging characterization establishes a comprehensive reference standard for the various modification techniques studied, each of which have applications in numerous fields.

## II. EXPERIMENTAL METHODS

Some of the following experimental methodology and results have been reported previously by our group [1], however this section elaborates upon the complete experimental procedure and associated data set. As such, the various techniques are color-coordinated across tables and figures to facilitate easier cross-referencing.

### A. Chemicals and Materials

The following materials were used as received: Sylgard 184 silicone elastomer (Dow Corning), 4-(1,1,3,3-Tetramethylbutyl)phenyl-polyethylene glycol (Triton X-100) (Sigma), Polyethylene glycol sorbitan (Tween 20) (Sigma), Pluronic F-127 (Sigma), 2-hydroxyethyl methacrylate 99+% (HEMA) (Sigma), sodium dodecyl sulfate 99+% (SDS) (Sigma), Poly(dimethylsiloxane-b-ethylene oxide) 25:75 (PDMS-b-PEO) (Polysciences), polyalkyleneoxide modified heptamethyltrisiloxane/allyloxypolyethyleneglycol methyl ether (Silwet L-77) (Fisher), benzophenone 99+% (Sigma), xylenes (Fisher), acetone (Fisher), toluene (Fisher), ethanol (Warner Graham), and low density polyethylene petri dishes.

### B. PDMS Fabrication

Six distinct modifications, described in subsequent sections, were investigated alongside pristine, unmodified Sylgard 184 PDMS to characterize their relative hydrophilic behavior. All PDMS samples were cured in an oven at 65 °C for 4 hours, unless otherwise noted.

### C. PDMS Modifications

*1) Surfactant Embedding*

Selected nonionic surfactants were added to the PDMS prepolymer mixture before curing. These amphiphilic molecules hydrophilize PDMS by molecular rearrangement of polar groups to the surface upon introduction of water. Chosen surfactants and concentrations (wt. %) are listed in Table I.

*2) Common Additives*

The commonly used additives benzophenone and carbon black were added to a Silwet L-77 sample to determine how molecular modifications might affect surfactant embedding hydrophilization.

*3) Extracted PDMS Oxidation*

An early method for fabricating long-term hydrophilic samples, introduced by the Whitesides group [25], relies on the extraction of un-crosslinked low-molecular weight oligomers from the bulk of a cured PDMS substrate. This technique was adapted using three pairs of consecutive solvent soaks: 1 hr in 200 mL xylenes (repeat), 1 hr in 200 mL toluene (repeat), and 1 hr in 200 mL acetone (repeat). After 4 hr drying at 65 °C, surfaces were treated with $O_2$ plasma at 100 W for 30s.

*4) Substrate-mediated Microroughening (Acetone Etch)*

Zhang et al. recently developed a method of roughening PDMS microfluidic channels for the purpose of isolating and recovering circulating tumor cells [26]. In adapting this method for our application, petri dishes were etched in 1:1 Acetone:EtOH without stirring for 150 and 300 s, after which, a pre-mixed, unadulterated PDMS batch was poured into the dishes. After degassing and curing, the samples were peeled from the etched dishes and treated with $O_2$ plasma at 100 W for 30 s in a PlasmaEtch PE25-JW plasma cleaner.

*5) RIE Surface Nanostructuring*

Select pristine, cured PDMS surfaces were structured with ion plasmas, adapted from literature [6], [27], using an Oxford PlasmaLab 80 with recipes outlined in Table II. One structuring method employed a single plasma etch step (recipes 1-2), whereas another utilized two etch steps (recipes 3-4).

*6) Polymer Grafting*

Select pristine PDMS samples were modified via surface grafting using hydrophilic monomers [6]. Specifically, after curing the PDMS, the surfaces were treated with $O_2$ plasma for 30 s at 100 W. Subsequently, ~0.5 mL HEMA was spin-coated at 1500 rpm for 15 s, then a second plasma treatment for either 150 s or 300 s was performed. HEMA hydrophilization may be attributed to monomeric carbonyl groups polarizing PDMS surfaces.

*7) Oxide Sputtering*

DC reactive magnetron sputtering of $Al_2O_3$ and $TiO_2$ was carried out in an AJA International ATC 2400 Sputtering System to deposit layers of 5 nm, 10 nm, and 25 nm of each material. A set of samples comprising a prepared PDMS and glass substrates, each used for measurement analyses, were loaded into a secondary chamber for 10 minutes to off-gas prior to introduction in the sputtering chamber. $TiO_2$ was sputtered with 2.53 sccm $O_2$ and an Ar flow rate of 30.0 sccm, which was optimized to produce stochiometric $TiO_2$ at the transition from metallic to compound deposition mode as described by Tavares et al. [28]. $Al_2O_3$ was sputtered with 3.44 sccm $O_2$ and an Ar flow rate of 30.0 sccm, at the same metallic to compound transition. Depositions for each material were performed at 200 W (DC) from a 2-inch target at a working distance of 8 cm in confocal configuration. Base pressure for sputtering deposition ranged between $1.33-2.33 \times 10^{-5}$ Pa. The working pressure during sputtering was 0.33 Pa for both $Al_2O_3$ and $TiO_2$ depositions.

*D. Characterization Techniques*

*1) Contact Angle Goniometry*

Images of DI water droplets on each surface were recorded using a Basler acA2500-60um camera with a 6X close focus zoom lens (Edmund). Images were analyzed using ImageJ to extract static contact angles. Specifically, "contact_angle.jar" [29] was used for circle fitting to contact angles < 20° and "DropSnake" [30] was used for active contour polynomial fitting for contact angles > 20°. Humidity was maintained between 15-20% in order to standardize measurements recorded over time. For long-term stability characterization, contact angles were measured over a 30-day period.

*2) Diffusivity Model*

Utilizing short-time contact angle measurements, a novel model was developed and data fitting in Python was utilized in order to extract molecular diffusion coefficients for each surfactant through bulk PDMS at various concentrations.

*3) Surface Topographical Analysis*

Both pristine and roughened PDMS surfaces were characterized using an NTEGRA Prima AFM. Images of size 5 x 5 μm and resolution 512 x 512 pixel were obtained using tapping mode, then analyzed to compare respective RMS roughness values.

*4) Surface Functionality Analysis*

In order to determine the surface functional groups, a selection of samples ($O_2$ and $SF_6$ plasma treated, HEMA grafted, PDMS-b-PEO embedded, and pristine PDMS) were analyzed using a Varian 3100 FTIR/ATR spectroscopy system with a ZnSe crystal. So as to identify functional group changes introduced by modifications, a pristine PDMS spectrum was subtracted from each modified PDMS spectrum to yield differential spectra.

### III. RESULTS & DISCUSSION

*A. Surfactant Embedding*

When amphiphilic surfactants are added to a PDMS mixture, water introduced on the surface promotes molecular diffusion of polar groups to the surface, increasing sample hydrophilicity. The kinetics of the transforming contact angle can be described by the Ward-Tordai equation (1), which gives the time-dependent surface excess concentration ($c_s$) of surfactant,

$$c_s(t) = 2c_0\sqrt{Dt/\pi} - 2\sqrt{D/\pi}\int_0^{\sqrt{t}} c_{ss} d(\sqrt{t-\tau}) \quad (1)$$

where $c_0$ is the bulk surfactant concentration, $D$ is the effective bulk diffusion coefficient, $\tau$ is a dummy variable, and $c_{ss}$ is the sub-surface concentration [12], [31]-[33]. For short times, the second term, related to the backwards diffusion from the surface back into the bulk, is negligible. Thus, the surface excess concentration becomes

$$c_s(t) = 2c_0\sqrt{Dt/\pi} \quad (2)$$

If one assumes an ideal gas diffusion and lattice adsorption process, the equation of state can be derived from the Gibbs energy of two bulk phases separated by a surface interface. The resulting Gibbs adsorption equation at the interface is

$$\sum_i n_i^s d\mu_i + A\, d\gamma = 0 \quad (3)$$

where $n_i$ is the number of adsorbed molecules of type $i$ on an area $A$ with chemical potential $\mu_i$ and surface tension $\gamma$. Defining the phase-dividing surface as having no surface excess of solvent (PDMS), surfactant surface excess ($\Gamma$) can be related to surface tension as

$$d\gamma = -RT\Gamma \frac{dc}{c} \quad (4)$$

where $R$ is the gas constant, $T$ is the temperature. Combining (2) and (4) and integrating, we can compare surface tension vs. time for short times as

$$\gamma(t) - \gamma_0 = -2c_0 RT\sqrt{Dt/\pi} \quad (5)$$

Lastly, inserting $\gamma(t)$ into the Young-Dupre equation [34] yields

$$\cos\theta = \frac{2c_0 RT}{\gamma_{LV}}\sqrt{D/\pi}\sqrt{t} + \frac{\gamma_{SL_0} - \gamma_{SV}}{\gamma_{LV}} \quad (6)$$

where $\gamma_{LV}$, $\gamma_{SV}$, and $\gamma_{SL_0}$ are the surface tensions of the liquid-vapor ($H_2O$-Air), solid-vapor (PDMS-Air), and initial solid-liquid (PDMS-$H_2O$) interfaces. Therefore, plotting $\cos\theta$ vs. $\sqrt{t}$ produces a linear graph, the slope of which is used to extract the surfactant diffusivity, $D$. Experimental plots of varying concentrations of Silwet L-77 are shown in Fig. 1(a), and corresponding diffusivity values for each surfactant plotted in Fig. 2.

This approach was considered, but not pursued, by Holczer et al., in favor of a simpler "quasi reverse-proportional" relation between surface excess and contact angle, from which diffusivities were not able to be quantitatively studied [12]. Fig. 1(a) demonstrates the various temporal regimes present in the diffusivity experiment. At low concentrations, there is a delay in the effect of wetting on the surface. This delay has been noted or observed in some studies, but not examined in detail [8], [9], [12]. The origin of the delay is unknown, though we hypothesize that it arises through either contact line pinning (a phenomenon caused by rough or chemically inhomogeneous surfaces) or a radial dependence of the chemical potential for interface-adsorbing molecules. The latter could result in delayed surfactant adsorption at the contact line. Consequently, an experiment was carried out to seek a dependence of the delay time on droplet volume. No correlation between volume and delay time was found, however, additional studies are required to confirm this observation and further characterize the phenomenon. Noting the lack of a delay time volume dependence, we attribute the phenomenon to contact line pinning. The decrease in delay time with increased concentration is compatible with this theory, as an increased adsorptive driving force more quickly overcomes the energetic barrier for wetting.

The second temporal regime represents the linear regime of forward diffusion, to which linear models were fitted and diffusivities extracted. The final regime is where our linear approximation breaks down due to backwards diffusion and possible interaction between surface adsorbed molecules.

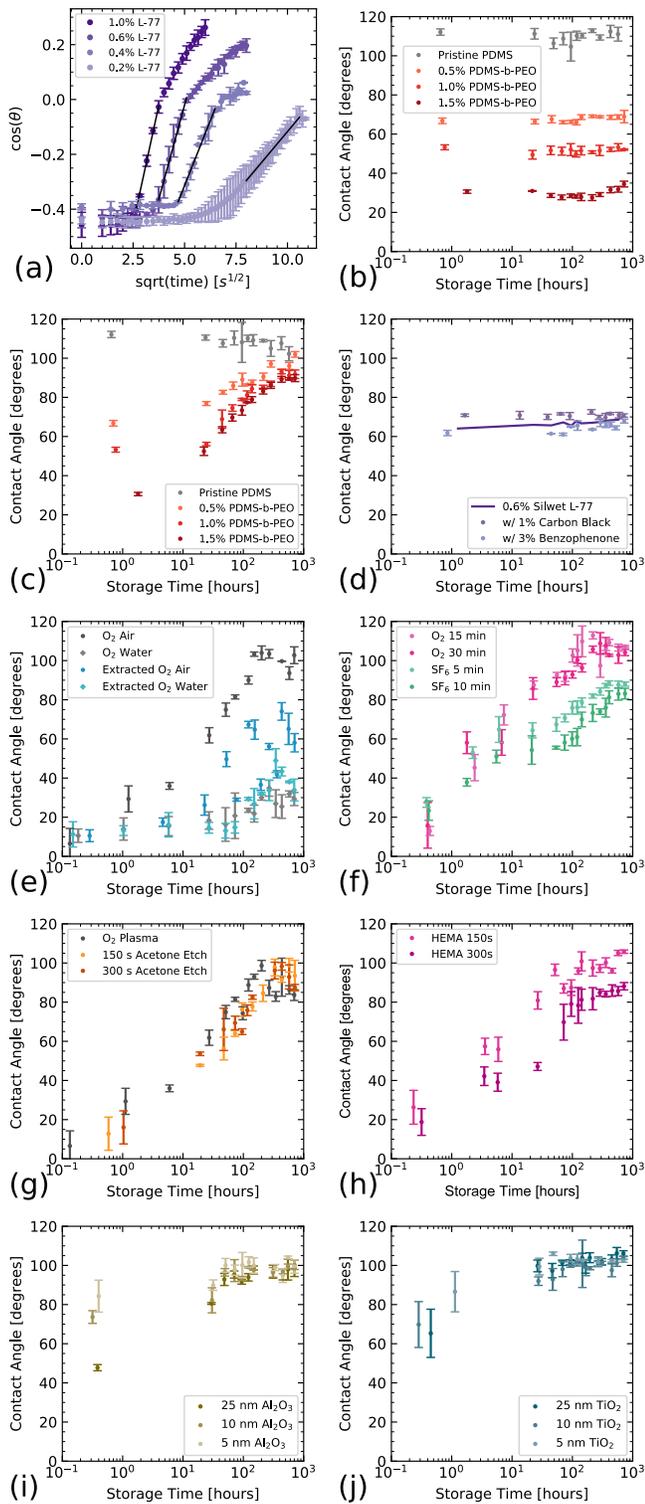

Fig. 1. Linear fitting of the cosine of contact angles vs. sqrt(time) (a). Longitudinally evolving hydrophilicity can be characterized by contact angles for various methods of hydrophilization. Methods of hydrophilization include nonionic surfactant addition in the bulk—stored in air (b), water (c), and with additives (d)—plasma activation (e), reactive ion etching (f), substrate etching (g), polymer grafting (h), sputtering of $Al_2O_3$ (i) and $TiO_2$ (j) oxide layers. These samples were stored in air unless otherwise noted.

As demonstrated in Silwet L-77 samples, low concentration (0.2-0.6 wt.%) samples result in consistent diffusivity values,

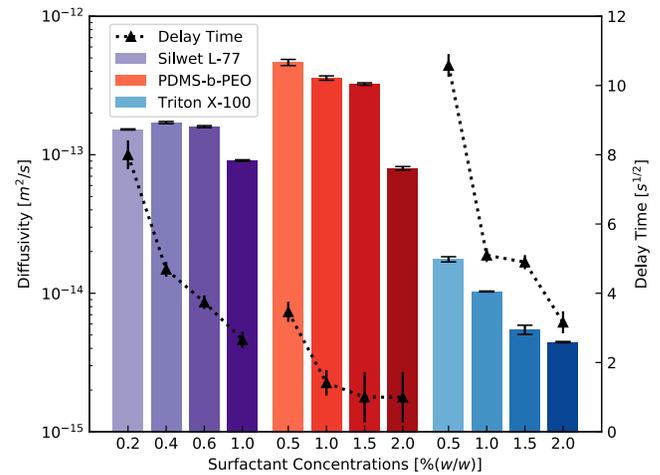

Fig. 2. The diffusivities of Silwet L-77 (purple), PDMS-b-PEO (red), and Triton X-100 (blue) as a function of concentration. The length of the non-evolution temporal regime attributed to contact line pinning is plotted as "delay time," including SD bars.

averaging $\mathbf{1.56 \pm 0.04 \times 10^{-13}\ m^2/s}$ across varying concentrations. In contrast, when concentration is increased (≥1%), the linear model breaks down, resulting in lower measured diffusivities. It is unclear whether this is due to localized micellar formation within the PDMS matrix, surfactant self-interactions, or a change in the diffusion behavior of surfactants (acting as a non-ideal gas). Further work will be carried out to further characterize surfactant behavior in PDMS and other polymer matrices.

Of the three surfactants, the fastest measured diffusivity was PDMS-b-PEO at $\mathbf{4.6 \pm 0.2 \times 10^{-13}\ m^2/s}$ (for 0.5 wt.%). Triton X-100 was the slowest diffuser at $\mathbf{1.76 \pm 0.08 \times 10^{-14}\ m^2/s}$ (for 0.5 wt.%). It has been shown that diffusivity has a power dependence on the molecular weight of the diffusing species [35], but with a smaller effect in PDMS as compared to other polymers [36]. Other parameters describing size, shape, and flexibility of the diffusing molecules may be of more use, and will be interesting to study for surfactants in PDMS [37], [38]. This technique presents a simple method for studying molecular diffusivities in polymer networks, which are of great importance for applications in drug delivery, degradable polymers, the development of artificial organs, and polymeric molecular separation membranes for the environmental industry.

When considering device shelf-life, especially in the context of commercializing a microfluidic medical product, long-term stability becomes a critical parameter. Fig. 1(b) and Table III illustrate the high stability of contact angles over a timescale of 30 days for three PDMS-b-PEO concentrations in comparison with pristine PDMS while being stored in air. The measured contact angles correspond to the recorded values after 3 minutes of contact with water. PDMS-b-PEO exhibited decreased effectiveness when stored in water due to leaching of surfactant over time (Fig. 1(c)). This would affect experiments using surfactant-modified PDMS microfluidic platforms, such as cell studies. Both Triton X-100 and Silwet L-77 demonstrated similar patterns of stability.

Interestingly, at higher concentrations of each surfactant, samples became increasingly opaque. This can cause issues for microfluidic devices where high visible-wavelength

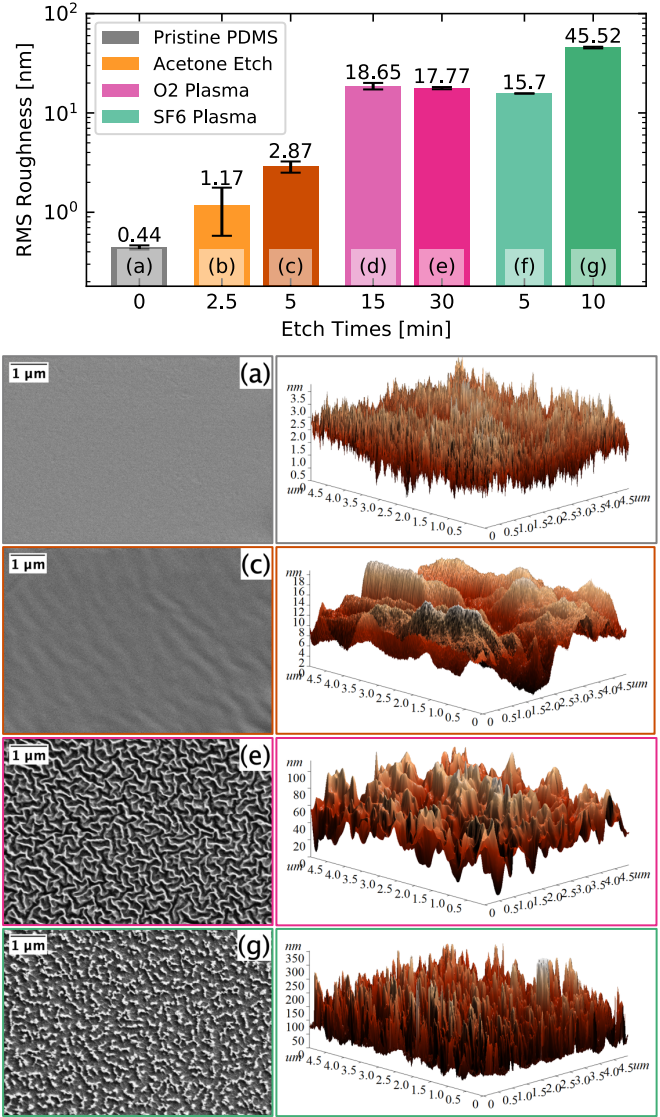

Fig. 3. (top) Average RMS roughness of etched surfaces with SD bars. (bottom) Associated SEM and 3-D rendered AFM plots.

TABLE III
INITIAL AND FINAL CONTACT ANGLES

| MODIFICATION PROCESS | STATIC CONTACT ANGLES ± S.D. [DEG] | | |
|---|---|---|---|
| | INITIAL θ | FINAL θ (AIR) | FINAL θ (WATER) |
| PEO-b-PDMS 1.5% | 30.6 ± 0.2 | 34.5 ± 1 | 89.8 ± 2 |
| Silwet 0.6% | 65.1 ± 0.4 | 70.6 ± 0.5 | 93.4 ± 0.5 |
| Triton 3% | 76.0 ± 0.8 | 78.3 ± 0.4 | 85.1 ± 0.9 |
| $O_2$ Plasma | 9.8 ± 3 | 102.7 ± 4 | 29.3 ± 3 |
| Extracted / $O_2$ Plasma | 10.6 ± 3 | 58.0 ± 4 | 33.9 ± 6 |
| RIE $O_2$ 15 min | 13.0 ± 2 | 107.2 ± 1 | - |
| RIE $SF_6$ 10 min / $O_2$ 30 s | 23.4 ± 5 | 83.1 ± 3 | - |
| HEMA / 300 s $O_2$ Plasma | 18.8 ± 7 | 88.1 ± 2 | - |
| Acetone 300 s | 16.0 ± 8 | 87.5 ± 2 | - |
| $Al_2O_3$ 25 nm | 47.8 ± 2 | 98.6 ± 4 | - |
| $TiO_2$ 25 nm | 65.3 ± 10 | 106.1 ± 1 | - |

PDMS samples when stored in water, as polar interaction with the aqueous environment remained dominant on the surface and consequently, molecular rearrangement of surface silanol moieties was energetically unfavorable. As the extraction causes swelling of the PDMS substrate, it must be carried out before post-processing, making this procedure incompatible with certain fabrication processes—for example, if electronics have been embedded in the PDMS.

*D. RIE Surface Nanostrucutring*

In the case of homogeneous (Wenzel state) wetting, surface roughness correlates with the roughness ratio, $r$, as

$$\cos(\theta_m) = r \cos(\theta_Y) \qquad (7)$$

where $\theta_m$ is the measured static CA and $\theta_Y$ is the Young CA. Stable surface roughness can be generated through anisotropic reactive ion etching. As has been demonstrated by the Gogolides group and confirmed here, $SF_6$ anisotropically etches nanocolumns into PDMS surfaces [27]. Whereas $O_2$ plasma etched samples saw negligible change in roughness for doubled etch time from 15 to 30 mins, $SF_6$ plasma etched samples showed a 190% increase in RMS roughness for doubled etch time from 5 to 10 mins, as seen in Fig. 3. These differences correlated with wettability and hydrophobic recovery, as $SF_6$ etched and oxidized samples had substantially delayed recovery timescale and lower final contact angle than shorter etched, while insignificant change was demonstrated by increased $O_2$ etch time (Fig. 1(f)). These changes are described by the replacement of surface methyl groups with hydroxyl and S-F groups (Table IV, Fig. 4). The success of the $SF_6/O_2$ sequential treatment is evident as the surface tension plateaus at a slightly hydrophilic contact angle of 83° after one month. This hydrophilic retention is attributed to two factors: (1) increased surface area of the roughened samples leads to higher hydroxylation density, and (2) the post-plasma stiff oxidized, nanostructured surface layer retards rearrangement of low-molecular weight hydrophobic moieties to the surface. These effects have been noted previously in the literature [27]. Additionally, XPS spectra have identified alumina and silica on plasma-etched polymer surfaces, which may further delay the hydrophobic recovery times [39]. This surface roughness can be multipurposed for applications such as improving cell growth and adhesion within microchannels [40].

*E. Substrate-mediated Microroughening (Acetone Etch)*

Wet etching of petri dishes for PDMS molding allowed for transparency is of importance. Additionally, it may provide an explanation to the failure of the presented model at high concentrations: if surfactant molecules are aggregating through micellization or similar phenomena, it may present as slower diffusivity and increased scattering of incident light.

*B. Common Additives*

Carbon black and benzophenone had negligible effects on the surfactant hydrophilization of PDMS surfaces (Fig. 1(d)).

*C. Extracted PDMS Oxidation*

When PDMS is immersed in a high swelling solvent, low-molecular weight oligomers are able to leach out from the bulk. As a result, when the surface is hydrophilized with an oxygen plasma, these oligomers are unable to replace polar silanol groups on the surface [25]. This allowed for significantly extended hydrophilic stability of samples stored in air with a final contact angle of 61°. This can be seen in Fig. 1(e), compared with extracted samples stored in water, and non-extracted samples in water and air. Extracted samples showed insignificant improvement over non-extracted plasma-treated

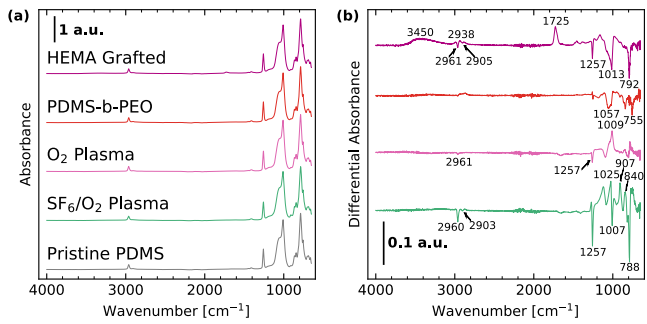

Fig. 4. (a) FTIR-ATR of various samples; (b) differential spectra by subtracting pristine PDMS spectrum.

TABLE IV
FTIR-ATR DIFFERENTIAL SPECTRA PEAKS

| SAMPLE | GROUP | WAVENUMBER [cm$^{-1}$] | +/- |
|---|---|---|---|
| HEMA | OH | 3450 | + |
|  | HEMA CH$_3$ | 2938 | + |
|  | C=O | 1725 | + |
|  | CH$_3$ | 2961, 2905, 1256 | - |
|  | Si-O-Si | 1013 | - |
|  | Si-(CH$_3$)$_x$ | 792 | - |
| PDMS-b-PEO | Si-O-Si | 1057 | - |
|  | Si-(CH$_3$)$_x$ | 755 | - |
| O$_2$ Plasma | CH$_3$ | 2961, 1257 | - |
|  | Si-O-Si | 1009 | + |
|  | Si-(CH$_3$)$_x$ | 803 | - |
| SF$_6$/O$_2$ Plasma | CH$_3$ | 2960, 2903, 1257 | - |
|  | Si-O-Si | 1007 | - |
|  | Si-(CH$_3$)$_x$ | 788 | - |
|  | Si-F$_x$ | 1025, 907, 840, 805 | + |

Identification of peaks (+) and troughs (-) in the differential absorbance of corresponding spectra. *Notes:* (1) Some differential peaks appear shifted due to reduction in nearby bands, e.g. SF$_6$ 788 cm$^{-1}$ Si-(CH$_3$)$_2$, and vice versa. (2) Sharp Si-OH peak at ~3750 cm$^{-1}$ is not visible as it is orders of magnitude smaller in intensity compared with the recorded spectra [6].

microroughening of the PDMS surfaces. The RMS roughness increased from 0.44 nm for a pristine petri dish, to 2.87 nm following a 5 min acetone etch. After subsequent plasma treatment, samples demonstrated negligible delay in hydrophobic recovery as compared to standard plasma treated samples, as both 150s and 300s etched samples recovered to a hydrophobic state within 8 days as seen in Fig. 1(g). This is likely because the degree of roughening was over an order of magnitude smaller than is possible with RIE etching, offering very minor increases in the hydroxyl group surface density. Additionally, the surface would not have accrued the same stiff oxide layer or silica/alumina seen in RIE-roughened samples. This method does, however, form smooth ridges, which may be useful for cell isolation and attachment, and the technique requires no post-processing, making it more compatible with various applications [26].

*F. Polymer Grafting*

Grafting of HEMA onto PDMS surfaces delayed and impeded hydrophobic recovery, with increasing effectiveness for longer plasma treatment times. This process introduces a more permanent hydrophilic state, but still displays recovery patterns characteristic of PDMS as seen in Fig. 1(h). A final contact angle of 88° suggests the viability of the process for a short-term device application, but with less promise for use in medical devices. FTIR-ATR spectral analysis confirmed HEMA presence on the PDMS surface as seen in Fig 4 and

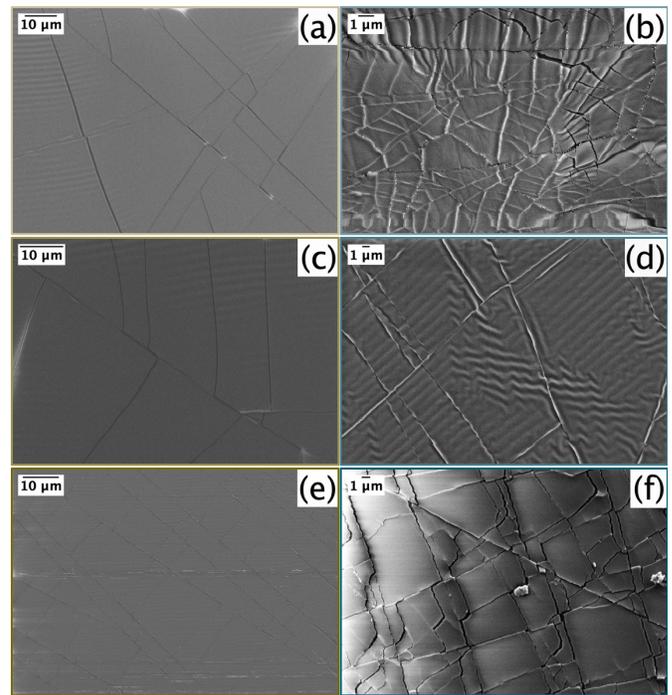

Fig. 5. SEM images of oxide layers deposited on PDMS. Al$_2$O$_3$ oxides (*left*) demonstrated less surface cracking for all thicknesses—5 nm (*top*), 10 nm (*middle*), and 25 nm (*bottom*)—when compared with TiO$_2$ layers of equal thicknesses (*right*). This difference correlates with an increased hydrophobic recovery timescale in Al$_2$O$_3$. Imaged at 5 kV accelerating voltage.

Table IV. Pairing of an extraction step with polymer grafting may present a promising method of improving biocompatibility and hydrophilicity, however, the grafting step may place limitations on device applications.

*G. Oxide Sputtering*

As deposited films are expected to be amorphous per the literature [41], [42].

Both sputtered Al$_2$O$_3$ and TiO$_2$ increased the hydrophilicity of PDMS surfaces. Initial hydrophilization increased with oxide thickness, and Al$_2$O$_3$ samples recovered to a hydrophobic state more slowly than TiO$_2$. However, all materials and thicknesses were unsuccessful at maintaining hydrophilicity for over 2 days, and each reached a steady hydrophobic state and contact angle by the 5$^{th}$ day. This is attributed to adsorption of organic molecules and the well know hydration process of oxide films as analyzed by various groups [43]–[46]. Upon hydration, similar films have been shown to exhibit decreased surface energy (increased hydrophobicity). Furthermore, cracking in the oxide films from flexion of the PDMS substrate results in gaps in the films as seen with SEM imaging in Fig. 5. Although hydrophilicity is unstable, we present here a recipe for sputtering of Al$_2$O$_3$ on a flexible substrate, which may be of use for future applications in flexible electronics or sensors, and acts to successfully hydrophilize the surface in the short-term.

IV. CONCLUSION

PDMS has found much success in medical device fabrication due to its low cost and characteristic properties, but is intrinsically hydrophobic. To render its surface compatible with use in pumpless microfluidics, many methods of surface hydrophilization have been formulated. We studied the hydrophobic recovery of an array of modifications, utilizing contact angles to characterize surface energy over a 30-day study, and FTIR-ATR, SEM, and AFM to characterize the surface properties. Low-molecular weight oligomer extraction using high-swelling solvents delayed and inhibited hydrophobic recovery by preventing molecular rearrangement of $O_2$ plasma-induced silanol groups on the surface, but is incompatible with some post-processed chips. Multiple roughening techniques demonstrated the utility of increased surface roughness on hydrophilic nature after silanolization, but still demonstrated similar hydrophobic recovery patterns. Metal oxides initially hydrophilized the surface, but cracking of the oxide layer caused failure of this technique. Addition of nonionic surfactants to the PDMS mixture yielded stable samples for air storage over 30 days, however requires contact time for molecular diffusion to the surface so as to induce hydrophilicity. We introduced a model for calculating surfactant diffusivities through a polymer matrix. PDMS-b-PEO was identified as the optimal surfactant for long-term device stability in air storage due to its long-term stability and fast diffusivity in PDMS. Other surfactants of similar chemical structure may be identified as superior, and further work is being conducted to characterize surfactant diffusion in polymer networks and its compatibility with device fabrication.

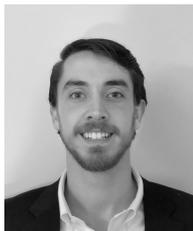

**Daniel J. O'Brien** was born in New York in 1995. He received a B.A. in physics from the College of the Holy Cross in 2017, a M.S. in physics from Georgetown University in 2019, and is now a Ph.D. candidate at the latter.

In 2015, he was granted an NSF-REU award for research in the Paranjape Nanotech and Biomedical Engineering lab at Georgetown. Since then, his research has expanded into micro- and nanofabrication, polymer science, and medical biotechnological applications.

Daniel has been inducted into *Sigma Pi Sigma*, and named both the Forster Family Scholar and Endowment Fellow by the ARCS-MWC.

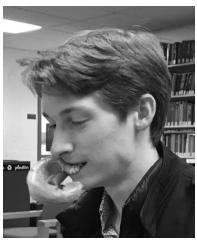

**Andrew J.H. Sedlack** was born in Stamford, CT in 1998. He received a B.S. in physics from Georgetown University in 2020. From 2018 to 2020 he had the privilege of working in the Paranjape Lab at Georgetown on polymer science and microfabrication.

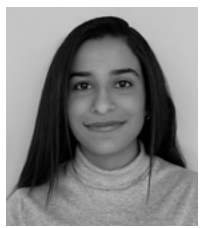

**Pia Bhatia** was born in New York in 1999. She is an undergraduate student at Georgetown University ('21), completing a B.S. in physics. Through her involvement with the Paranjape Group and GNuLab at Georgetown, she has developed an interest in micro- and nanofabrication research. In 2020, she was awarded the Hichwa Family Undergraduate Research Fellowship to further support her research with the Paranjape Lab. Following graduation, Pia plans to pursue a Ph.D. in physics.

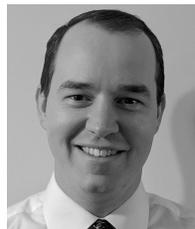

**Christopher J. Jensen** was born in Pennsylvania in 1991. He received his B.S. in physics in 2014 and his M.S. in applied physics in 2016 from Towson University. Additionally, he received an M.S. in physics from Georgetown in 2019, where he is currently pursuing a Ph.D.

Christopher's research interests include nanomagnetism, electric-field control of magnetism, novel materials growth techniques, and thin film characterization.

During his time at Towson and Georgetown University, Christopher has been a recipient of several graduate research and travel grants through the universities.

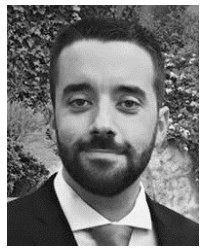

**Alberto Quintana** was born in Mataró, in the province of Barcelona (Spain), in 1987. He received the B.S. in Physics, a M.S in Materials Science and Nanotechnology at the Autonomus University of Barcelona (Spain) in 2014. He carried out his Ph.D. under the European Research Council Consolidator Grant *Spin-Porics*, at the same institution, graduating in 2018.

Dr. Quintana is currently a Post-doctoral Researcher at Georgetown University. His research interests include nanomagnetism, electric-field control of magnetism (i.e. charge accumulation effects, ferroelectric / ferromagnetic heterostructures and magneto-ionics), metallic/oxide nanofoams for filtering and catalytic properties and the development of novel materials/composites. He is author of 20 articles published in high-impact factor journals such as ACS Nano, Advanced Functional Materials, ACS Applied Materials and Interfaces or Small, and holds 1 patent.

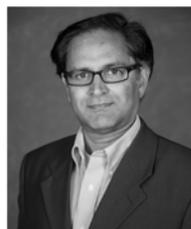

**Makarand (Mak) Paranjape** received his B.Sc. (1986), M.Sc. (1990), and Ph.D. (1993) from the University of Alberta, Edmonton, Canada in electrical engineering with a focus on semiconductor physics, microelectronics, and MEMS.

He is presently an Associate Professor in the Physics Department at Georgetown University, Washington, D.C., having joined the faculty in 1998, and is the current Director of the Georgetown Nanoscience and Microtechnology Lab (GNµLab), a multi-disciplinary cleanroom facility. In 2020, he began his term as Scholar-in-Residence at the US Food and Drug Administration (Silver Spring, MD), in the Office of Science and Engineering Laboratories, working on the detection of opioid biomarkers.

Prior to his faculty position, he was a post-doctoral researcher at Concordia University (Montreal, Canada), Simon Fraser University (Vancouver, Canada), and the University of California (Berkeley, CA). In 1995, he held an engineering consulting position for 3 years at the Istituto per la Ricerca Scientifica e Tecnologica (IRST) in Trento, Italy. He is an inventor of a unique biomedical technology for sensing human biomarkers using a non-invasive transdermal approach, and holds key intellectual property on the technology.

Dr. Paranjape serves as Associate Editor for *Biomedical Microdevices* and has served on review panels for the National Science Foundation (NSF), the National Institutes of Health (NIH), the Natural Sciences and Engineering Research Council of Canada (NSERC), US Food and Drug Administration (FDA), and the US Defense Threat Reduction Agency (DTRA). In 2011, Dr. Paranjape was awarded the *Georgetown College Dean's Teaching Award* for outstanding achievement in the teaching of undergraduate students. Dr. Paranjape's patent portfolio earned him the *Award for Outstanding Contribution in Innovation and Commercialization* from Georgetown University in 2013 and membership to the National Academy of Inventors in 2014.